\documentclass[aps,twocolumn,showpacs,floatfix, superscriptaddress, pra,reprint, footinbib, a4paper]{revtex4-2}

\usepackage{subfigure}
\usepackage{psfrag,graphicx}
\usepackage{pict2e}
\usepackage{dcolumn}
\usepackage{amsmath,amssymb,braket}
\usepackage{bm}
\usepackage{color}
\usepackage{latexsym}
\usepackage{epstopdf}
\usepackage{color}
\usepackage{comment}
\usepackage[english]{babel}
\UseRawInputEncoding
\usepackage{amsfonts}
\usepackage{bm}
\usepackage{natbib}
\usepackage{bbold}

\usepackage{enumerate}

\usepackage{tikz}

\definecolor{myblue}{rgb}{0.2,0.2,0.8}
\definecolor{myzard}{cmyk}{0,0,0.05,0}
\definecolor{mywhite}{rgb}{1,1,1}
\definecolor{mywhite}{rgb}{1,1,1}
\definecolor{myred}{rgb}{1,0.,0.3}
\usepackage[colorlinks=true,citecolor=myblue,linkcolor=myred]{hyperref}

\definecolor{darkgreen}{rgb}{0.0, 0.4, 0.26}

\definecolor{mygrey}{gray}{0.35}
\definecolor{myblue}{rgb}{0.2,0.2,0.8}
\definecolor{myzard}{cmyk}{0,0,0.05,0}
\definecolor{mywhite}{rgb}{1,1,1}
\definecolor{mywhite}{rgb}{1,1,1}
\definecolor{myred}{rgb}{1,0.,0.3}

\def\be{\begin{equation}}
\def\ee{\end{equation}}
\def\ba{\begin{align}}
\def\enda{\end{align}}
\def\bi{\begin{itemize}}
\def\ei{\end{itemize}}

\def\beq{\begin{equation}}
\def\beq{\begin{equation}}
\def\eeq{\end{equation}}

\begin{document}

\title{Programming long-range interactions in analog quantum simulators}

 \author{Cristian Tabares}
 \email{cristian.tabares@csic.es}
 \affiliation{Institute of Fundamental Physics IFF-CSIC, Calle Serrano 113b, 28006 Madrid, Spain}
\author{Alberto Mu$\tilde{\mathrm{n}}$oz de las Heras}
\affiliation{Departamento de F\'isica, Facultad de Ciencias Ambientales y Bioqu\'imica, Universidad de Castilla-La Mancha, 45004 Toledo, Spain}

\author{Jan T. Schneider}
 \affiliation{Institute of Fundamental Physics IFF-CSIC, Calle Serrano 113b, 28006 Madrid, Spain}
 \author{Alejandro Gonz\'alez-Tudela}
 \email{a.gonzalez.tudela@csic.es}
 \affiliation{Institute of Fundamental Physics IFF-CSIC, Calle Serrano 113b, 28006 Madrid, Spain}

\begin{abstract}
Long-range interactions are the source of many equilibrium and out-of-equilibrium quantum many-body phenomena. Analog simulators based on ionic, atomic, superconducting, and molecular systems provide a natural platform to obtain these interactions using vibration- and photon-mediated processes. Recent experimental advances, such as their integration in multi-mode cavities and waveguides, or the use of Raman-assisted transitions, enable dynamical control over both the strength and the spatial range of these interactions, thereby rendering them programmable. Here, we develop a hybrid classical-quantum toolbox that exploits this tunability to enhance many-body state preparation in analog simulators beyond fixed-connectivity architectures. Our approach is based on classical pre-compilation in homogeneous small systems, whose optimized parameters are extrapolated iteratively to larger system sizes, and then refined on the quantum hardware using noise-aware hybrid re-optimization and error-mitigation techniques. We benchmark this strategy across several fermionic, spin-$1/2$, and spin-1 models, demonstrating orders-of-magnitude improvements in fidelity and energy estimation for system sizes ranging from $10^2$ to $10^3$ particles. Finally, we show that the combination of such high-fidelity programmable state preparation techniques with tunable-range out-of-equilibrium dynamics enables controlled studies of many-body thermalization in regimes accessible to current experimental platforms. Our results establish programmable long-range interactions as a powerful resource for next-generation analog quantum simulators.
\end{abstract}

\maketitle

\section{Introduction}~\label{sec:introduction}

Long-range interactions play a central role in quantum many-body physics~\cite{Defenu2023}, leading to competing orders and exotic many-body phases~\cite{Leonard2017,Semeghini2021,Liu2022,Mann2025arXiv,hauke10c,Maik2012a,Arguello-Luengo2022}, while strongly impacting quenched and driven many-body dynamics~\cite{Neyenhuis2017,Sugimoto2022,Mattes2025,Nandkishore2017,Serbyn2021,Lerose2025,Hauke2014,Richerme2013}. However, their study with classical methods is hindered by the rapid growth of entanglement and correlations induced by long-range interactions~\cite{Hauke2013_spreadLR,Koffel2012}. A complementary approach is to engineer these long-range interacting models~\cite{Defenu2023,ArguelloLuengo2025} in platforms such as trapped ions~\cite{Blatt2012,Monroe2021}, Rydberg atoms~\cite{Chomaz2022,Browaeys2020}, superconducting circuits~\cite{majer07a,Zhang2022Simulator}, or molecules~\cite{Micheli2006}, which can emulate these systems as analog quantum simulators~\cite{cirac12a}. While these devices have provided insights into equilibrium~\cite{britton2012,islam13a,Chomaz2016} and dynamical~\cite{richerme14a,Jurcevic2014,Smith2016,zhang17a,Bernien2017,Choi2017} regimes of certain models, their analog nature restricts the accessible many-body phases and dynamics due to limited native Hamiltonian tunability and small energy gaps that hinder adiabatic state preparation in certain regions of the phase diagrams. These challenges limit both the scalability and flexibility of these simulators, motivating the search for other operational paradigms.

\begin{figure*}[tb]
    \centering
    \includegraphics[width=\linewidth]{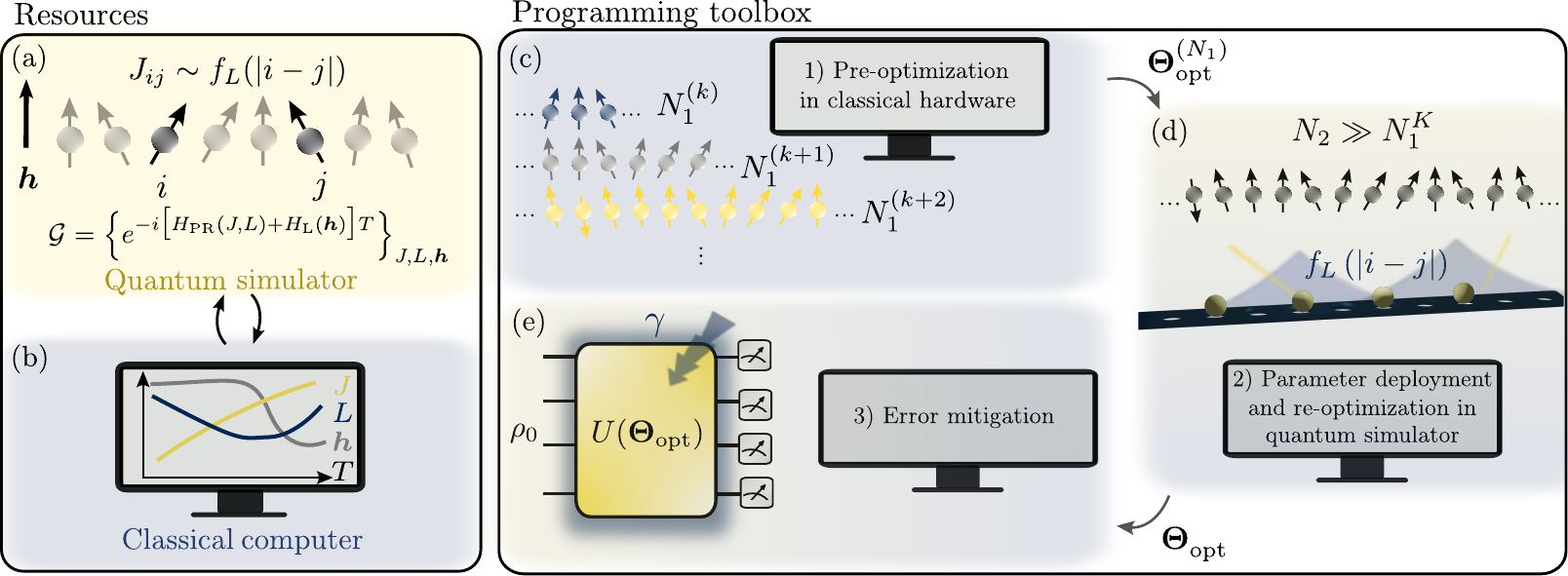}
    \caption{(a) Quantum resource: evolution under programmable-range (PR) Hamiltonians with on-site drivings $H_{\text{L}}(\bm{h})$ and tunable-range interactions $H_{\text{PR}}(J,L)$, whose spatial profile is set by $f_L(|i-j|)$. These operations define a gate set $\mathcal{G}$ used to construct parametrized quantum circuits. (b) Classical resource: pre- and post-processing techniques to calculate the optimal pulses to be implemented in the simulator using the hybrid workflow combining classical and quantum resources described in panels (c-e). (c) Pre-optimization in classical hardware, where circuit parameters are optimized in small systems $\bm{\Theta}_{\text{opt}}^{(N_1)}$ and iteratively extrapolated to larger sizes. (d) Deployment on quantum hardware with a hybrid optimization loop to refine parameters and correct calibration errors. (e) Error mitigation via, e.g., zero-noise extrapolation to accurately estimate observables.}
    \label{fig:intro}
\end{figure*}

Recent experimental advances are, nevertheless, expanding the capabilities of these analog devices. For instance, hybrid platforms combining atoms with cavities~\cite{ritsch13a,Mivehvar2021a,Vaidya2018}, waveguides~\cite{Chang2018,Sheremet2021,Gonzalez-Tudela2024}, or additional atomic species~\cite{Arguello-Luengo2019,Arguello-Luengo2021,Arguello_Luengo2022,DeSalvo2019,Edri2020} enable tunable spatial interaction profiles~\cite{ritsch13a,Mivehvar2021a,Vaidya2018,Chang2018,Sheremet2021,Gonzalez-Tudela2024,Arguello-Luengo2019,Arguello-Luengo2021,Arguello_Luengo2022,DeSalvo2019,Edri2020,douglas15a,Gonzalez-Tudela2015b,Hung2016,Gonzalez-Tudela2018,Gonzalez-Tudela2019a,Bello2019a,Leonforte2020b}, while Raman-assisted processes provide dynamical control over their strength and range~\cite{Blatt2012,Monroe2021,Hung2016,Periwal2021}. These capabilities open a new operational regime in which analog simulators can be used as programmable quantum devices~\cite{Daley2022}, implementing sequences of tunable entangling operations interleaved with on-site drivings. In this picture, the evolution can be viewed as a parametrized quantum circuit that is optimized for tasks such as ground-state preparation or controlled non-equilibrium dynamics beyond the native Hamiltonians. Standard approaches in this framework, such as the Variational Quantum Eigensolver (VQE)~\cite{Cerezo2021}, rely on hybrid classical-quantum optimization loops. However, fixed-connectivity circuit architectures---most  notably, those involving nearest-neighbors (NN) unitaries---often limit both their expressibility, i.e., the Hilbert space region they can access; and their trainability, i.e., their capability to find the optimal solution~\cite{LaroccaThanasilpWangSharmaBiamonteColesCincioMcCleanHolmesCerezo2025BarrenPlateausVQC}.
While programmable-range interactions have been suggested to mitigate these limitations in small systems for certain problems~\cite{Tabares2023,Lyu2023}, their use at larger scales, across different models, and under realistic experimental constraints remains an open question.

In this work, we address this challenge by introducing a hybrid classical--quantum framework, illustrated in Fig.~\ref{fig:intro}, that leverages the evolution of programmable-range (PR) quantum simulators [schematically depicted in Fig.~\ref{fig:intro}(a)] with classical pre- and post-processing techniques [as shown in Fig.~\ref{fig:intro}(b)]. The central ingredient is a classical pre-compilation strategy, see Fig.~\ref{fig:intro}(c), in which optimal circuit parameters are determined in small systems and iteratively extrapolated to larger sizes. Benchmarking this approach across a variety of homogeneous spin and fermionic models, we find that PR circuits yield orders-of-magnitude improvements over fixed-range architectures, enabling accurate state preparation for systems comprising up to $10^3$ modes. We further extend this strategy by incorporating zero-noise extrapolation~\cite{Temme2017} and a noise-aware hybrid re-optimization loop, see Figs.~\ref{fig:intro}(d-e), thereby constructing a complete pipeline for energy estimation and parameter refinement under realistic experimental constraints, including hardware noise, finite measurement budgets, and imperfect Hamiltonian control. Finally, we show that the combination of such high-fidelity programmable state preparation and tunable-range dynamics enables controlled studies of many-body thermalization~\cite{Neyenhuis2017,Sugimoto2022,Mattes2025}. As a concrete example, we prepare localized phases of the Aubry-Andr\'e-Harper (AAH) spin model~\cite{Harper1955,Aubry1980,Thouless1983} using PR circuits, and investigate their subsequent evolution under long- and short-range interacting Hamiltonians. Our results indicate that signatures of thermalization induced by integrability breaking through long-range interactions should be observable with system sizes accessible to current trapped-ion~\cite{Blatt2012,Monroe2021} and microwave waveguide QED setups~\cite{zhang2023}, opening new avenues for quantum simulators beyond simply preparing equilibrium states.

The rest of the manuscript is organized as follows: in Section~\ref{sec:Hamiltonians}, we introduce the PR interacting Hamiltonians that we use as the \emph{resource} for building the parametrized quantum circuits along the manuscript. In Section~\ref{sec:toolbox}, we explain the different methods in the optimization toolbox that we develop for exploiting PR simulators. In Section~\ref{sec:benchmark}, we systematically benchmark the scalability and generality of the toolbox approximating the ground states a variety of \emph{target} Hamiltonians, ranging from spin-$1/2$ and spin-$1$ models to fermionic ones. In Section~\ref{sec:dynamics}, we study the thermalization of many-body states prepared with this toolbox, and show numerical evidence that PR simulators can probe signatures of different thermalization regimes. Finally, in Section~\ref{sec:conclusions}, we summarize our findings and point to future work directions.

\section{Programmable Long-range Resource Architectures~\label{sec:Hamiltonians}}

Along this work, we consider one-dimensional PR simulators, whose dynamics is governed by Hamiltonians containing both on-site driving terms and interaction terms, see Fig.~\ref{fig:intro}(a). This setting captures a broad class of experimentally relevant platforms, including trapped-ion chains and waveguide QED, while allowing us to systematically investigate scalability to large systems.

The on-site driving term of the Hamiltonians we consider can be generally written as:
\begin{equation}~\label{eq:resource_SR}
    H_{\mathrm{L}} \left(\bm{h}\right) = \sum_{i} h_i O_{i}\,,
\end{equation}
where $i$ is the index running over the constituents of the system, $\bm{h}=(h_1,...h_N)$ is the vector containing the strength of each local driving, and $O_{i}$ are the on-site Hermitian operators associated to them, whose nature depends on the hardware considered. The number of independent parameters considered is typically not extensive in the system size, but only matches the size of a single unit cell instead. For instance, for homogeneous Hamiltonians we use only a single $h_i \equiv h$. 

On the other hand, the PR interaction terms we consider generally read:
\begin{equation}~\label{eq:resource_LR2}
    H_{\mathrm{PR}} \left(J,L\right) = J \sum_{i<j} f_L\left(\left|i-j\right|\right) O^{\dagger}_{i}O_{j}+\mathrm{H.c.}\,,
\end{equation}
where $J$ is the overall strength, and $f_L\left(\left|i-j\right|\right)$ the function determining the spatial decay of the interactions, whose range is parametrized by $L$. 

Within the programmable framework we adopt along this work, we assume that $J$, $\bm{h}$, and $L$  can be controlled dynamically. Therefore, we take $H_{\mathrm{L}} \left(\bm{h}\right) + H_{\mathrm{PR}} \left(J,L\right)$ as the \emph{resource} Hamiltonians to construct the parametrized circuits by concatenating unitary gates $\mathcal{G}$ generated by their dynamics as follows:
\begin{equation}~\label{eq:gate_set}
    \mathcal{G}(J,L,\bm{h},T) = \Big\{e^{-i\big[H_{\mathrm{PR}} \left(J,L\right) + H_{\mathrm{L}}\left(\bm{h}\right)\big]T}\Big\}\,.
\end{equation}
The challenge of programming the programmable-range quantum simulator is then reduced to finding the optimal parameters implementing a certain task, as shown in Fig.~\ref{fig:intro}(b). Also, note that this gate family, $\mathcal{G}$, interpolates continuously between purely on-site dynamics ($J\rightarrow 0$) and NN circuits ($f_{L\rightarrow0}(|i-j|)\rightarrow \delta_{|i-j|,1}$), which serves as the fixed-connectivity circuits against which we quantify the advantage of PR circuits.

In particular, we will consider several concrete realizations of Eq.~(\ref{eq:resource_LR2}) motivated by different platforms:
\begin{itemize} 

\item PR XX Hamiltonians reading:
\begin{equation}~\label{eq:H_exp_dec_xx}
    H_{\mathrm{PR}}^{(\mathrm{XX})} = J \sum_{i\neq j}f_L\left(\left|i-j\right|\right)\sigma_{i}^\dagger \sigma_{j}+\mathrm{H.c.}\,,
\end{equation}
with $\sigma_j^{\dagger}$ ($\sigma_j$) being a spin-$1/2$ raising (lowering) operator. This Hamiltonian appears naturally in waveguide and cavity QED setups~\cite{douglas15a,Gonzalez-Tudela2015b,Hung2016,Gonzalez-Tudela2018,Gonzalez-Tudela2019a,Bello2019a,Leonforte2020b} when emitters have their optical transition off-resonant from the photonic bands/cavity modes. Their spatial shape can feature both an exponential form, $f_L\left(\left|i-j\right|\right)=e^{-\left|x_i-x_j\right|/L}$, or a power-law shape, $f_L\left(\left|i-j\right|\right)=1/|x_i-x_j|^L$, if several Raman-assisted processes are considered~\cite{douglas15a,Gonzalez-Tudela2015b,Hung2016} or Dirac-like photons are used to mediate the interactions~\cite{Gonzalez-Tudela2018}.

\item PR Ising Hamiltonians of the form:
\begin{equation}~\label{eq:H_exp_dec_ising}
    H_{\mathrm{PR}}^{(\mathrm{I})} = J \sum_{i\neq j}f_L\left(\left|i-j\right|\right)\sigma_{i}^x \sigma_{j}^x\,,
\end{equation}
with $\sigma_j^{x} = (\sigma_j^+ + \sigma_j^-)/2$ being now a $x$-Pauli operator. This Hamiltonian with exponentially attenuated interactions can be engineered in waveguide and cavity QED setups using Raman-assisted processes in emitters with two optical transitions~\cite{douglas15a,Gonzalez-Tudela2015b,Hung2016}. The power-law version can also appear in trapped ion setups adding several Raman-assisted processes~\cite{Blatt2012,Monroe2021}.

\item PR spin-$1$ Blume-Capel (BC)~\cite{Blume66,Capel1966} Hamiltonians:
\begin{equation}~\label{eq:Ising_LR_spin1}
    H^{(\text{BC})}_{\mathrm{PR}} = -J\sum_{i\neq j} e^{-|i-j|/L} S_{i}^x S_{j}^x + D\sum_i (S_{i}^x)^2\,,
\end{equation}
with $S_{i}^x$ being spin-$1$ operators, similar to the spin-$1/2$ Ising ones but with an anisotropy term. Effective Hamiltonians as the ones in Eq.~\eqref{eq:Ising_LR_spin1} can appear in waveguide/cavity QED setups with emitters with three hyperfine levels~\cite{Tabares2022}.

\item PR fermionic quadratic Hamiltonians:
\begin{align}~\label{eq:H_resource_kitaev}
\begin{split}
    H_{\mathrm{PR}}^{(\mathrm{Kitaev})} = &J\sum_{i\neq j}f_L\left(|i-j|\right)\\&\left[\left(c^{\dagger}_{j} c_i +\mathrm{H.c.}\right)+ \left(c^{\dagger}_{j} c^{\dagger}_{i}+\mathrm{H.c.}\right)\right]\,,
\end{split}
\end{align}
which are the long-ranged version of the Kitaev chain model~\cite{Kitaev2001}. Although there are proposals to implement the long-range tunneling and pairing terms in current fermionic quantum processors~\cite{OBrien2018,Gonzalez-Cuadra_2023,Arguello-Luengo2019,Arguello-Luengo2020,Arguello-Luengo2021,Arguello_Luengo2022,Gkritsis2024,Li2026arxiv,Tabares2025,cavallar2025phasesensitivemeasurementsfermihubbardquantum}, the main motivation to introduce $H_{\mathrm{PR}}^{\mathrm{(Kitaev)}}$ as a resource Hamiltonian is that its non-interacting nature enables to solve the dynamics exactly using free-fermions~\cite{Surace2022a}. This will allow us to test the scalability of our optimization toolbox up to large system sizes, considering $\sim1000$ modes.
\end{itemize}

\section{Optimization Toolbox~\label{sec:toolbox}}

In this section, we introduce the optimization toolbox presented in Fig.~\ref{fig:intro}(c-e), which enables the scalable and experimentally realistic preparation of relevant many-body phases harnessing PR simulators. In Section~\ref{subsec:precompiled}, we first present a classical pre-compilation strategy designed to determine optimal circuit parameters in small systems and systematically extrapolate them to larger sizes. Building upon this scaling approach, in Section~\ref{subsec:zero} we show how to incorporate noise-mitigation techniques and hybrid re-optimization under imperfect Hamiltonian control to perform such optimization in realistic conditions.

\subsection{Classical Pre-Compilation and Scaling Strategy~\label{subsec:precompiled}}

Throughout this manuscript, we consider the preparation of the ground state of a target many-body Hamiltonian $H_{\mathrm{tar}}$ from an easy-to-prepare initial state, denoted by $\rho_0$. For that, we use exclusively the gates $\mathcal{G}$ of Eq.~\eqref{eq:gate_set}, which are generated by the resource Hamiltonians of Eqs.~\eqref{eq:resource_SR}--\eqref{eq:H_resource_kitaev}. To this end, we build parametrized quantum circuits by concatenating $D$ layers of unitary evolutions drawn from the gate family $\mathcal{G}$ introduced in Eq.~\eqref{eq:gate_set}, 
\begin{equation}
U(\bm{\Theta}) = \prod_{\ell=1}^{D} \mathcal{G}_{\ell}(J_\ell,L_\ell,\bm{h}_\ell,T_\ell),
\end{equation}
where $\bm{\Theta}$ collects all control parameters of the circuit. Acting on $\rho_0$, the circuit defines a parametrized state 
\begin{equation}
\rho(\bm{\Theta}) = U(\bm{\Theta})\rho_0 U^\dagger(\bm{\Theta}),
\end{equation}
whose energy with respect to the target Hamiltonian,
\begin{equation}~\label{eq:cost_function_VQE}
C_{\rho_0,H}(\bm{\Theta}) = \mathrm{Tr}\!\left[H_{\mathrm{tar}}\, \rho(\bm{\Theta})\right],
\end{equation}
serves as the cost function to be minimized. In standard VQE~\cite{Cerezo2021}, including implementations exploiting PR interactions~\cite{Tabares2023,Lyu2023}, this minimization is proposed to be performed within a hybrid classical--quantum loop. This means that the cost function is evaluated on the quantum device and then fed into the classical optimizer. While such schemes avoid the need of classically simulating the dynamics and can partially compensate for imperfect control, they suffer from trainability limitations as system size increases, reflected in increasingly flat optimization landscapes~\cite{LaroccaThanasilpWangSharmaBiamonteColesCincioMcCleanHolmesCerezo2025BarrenPlateausVQC} and the proliferation of sub-optimal local minima~\cite{AnschuetzKiani2022SwampedWithTraps}.

Inspired by recent warm-start strategies~\cite{EggerMarecekWoerner2021WarmStartingQuantumOptimization,Mele2022,RudolphMillerMotlaghChenAcharyaPerdomoOrtiz2023SynergisticPretraining,martin2026preoptimizationquantumcircuitsbarren,Puig2025,Farrell2024}, we instead adopt a scalable pre-compilation approach. We first perform the optimization fully classically for small system sizes $N_1$, obtaining a set of optimal parameters 
\begin{equation}
\bm{\Theta}_{\mathrm{opt}}^{(N_1)} = \mathrm{argmin}_{\bm{\Theta}}\, C_{\rho_0,H}(\bm{\Theta}),
\end{equation}
and iteratively extrapolate these to larger sizes, as schematically depicted in Fig.~\ref{fig:intro}(c). For the class of PR circuits and homogeneous target models considered here, we observe that the optimal interaction ranges and coupling strengths display smooth scaling behavior with system size, enabling systematic parameter transfer across different training sizes $N_1$. As we show in Section~\ref{sec:benchmark}, this iterative warm-start strategy allows us to prepare high-fidelity ground states for systems substantially larger than those used in the initial optimization.  This advantage can be intuitively understood from the ability of long-range interactions to generate correlations across the system faster than their NN counterparts, thereby reducing the circuit depth required to approximate highly entangled states. This is clear for power-law decaying interactions, in which the group velocity of correlation spreading diverges in some regimes~\cite{Schneider2021}. On the other hand, although exponentially-decaying interactions may still spread correlations ballistically like NN ones do~\cite{Cevolani2018}, one may also expect a parametric advantage arising from their use that leads to faster implementations of the circuits in hardware.

\subsection{Noise Mitigation and Hybrid Re-Optimization Under Imperfect Control~\label{subsec:zero}}

The pre-compiled extrapolation strategy described above assumes perfect control over the Hamiltonian parameters and noiseless circuits. In realistic experimental scenarios, however, both calibration errors and decoherence are unavoidable. As a consequence, the optimal parameters obtained under ideal conditions may no longer be optimal in the presence of noise. To address this issue, we complement the pre-compilation strategy with two additional techniques: a hybrid re-optimization on the quantum hardware and quantum error mitigation using zero-noise extrapolation (ZNE), which are depicted in Figs.~\ref{fig:intro}(d) and~(e), respectively.

The first source of experimental imperfection arises from parameter miscalibration. In practice, each optimized parameter may be implemented with small deviations, which we model as Gaussian fluctuations,
\begin{equation}~\label{eq:noise_params}
\bm{\Theta}_{\mathrm{opt}} \rightarrow \tilde{\bm{\Theta}}_{\mathrm{opt}}
= \bm{\Theta}_{\mathrm{opt}} + \bm{\varepsilon},
\qquad
\varepsilon_i \sim \mathcal{N}(0,\sigma^2)\,.
\end{equation}
Such deviations degrade the quality of the prepared state, even when the pre-compilation procedure yields high-fidelity parameters in the ideal setting. To compensate for these calibration errors, we propose to perform a hybrid classical-quantum re-optimization initialized at the pre-compiled parameter set. This refinement step allows the optimizer to adapt to hardware-specific imperfections while benefiting from a favorable starting point in parameter space, thereby mitigating the trainability issues encountered in large-scale variational optimizations.

On the other hand, for quantum error mitigation, we choose ZNE because of its natural implementation in analog simulators. This error mitigation scheme assumes the presence of a controllable noise rate $\gamma$ in the simulator. In analog platforms, this can be achieved, for instance, by rescaling the Hamiltonian strengths or modifying the total physical evolution time, effectively amplifying the impact of noise while preserving the underlying dynamics~\cite{QEM_ANALOG_1,steckmann2025errormitigationshottoshotfluctuations,guo2025mitigating}. In the presence of such noise, the cost function acquires an explicit dependence on $\gamma$, which we denote as $C_{\rho_0,H}(\bm{\Theta},\gamma)$. For sufficiently small noise rates, this dependence can be expanded around $\gamma = 0$ as
\begin{equation}
\begin{split}
C_{\rho_0,H}(\bm{\Theta},\gamma)
&= C_{\rho_0,H}(\bm{\Theta},0)
+ \frac{\partial C_{\rho_0,H}}{\partial \gamma}
\Big|_{\gamma=0}\gamma
+ O(\gamma^2)\,.
\end{split}
\end{equation}
By measuring the cost function at several amplified noise values and fitting its $\gamma$-dependence, one can extrapolate the estimate corresponding to the noiseless limit~\cite{Temme2017,Cai2023}.

\section{Benchmarking across target models~\label{sec:benchmark}}

In this section, we demonstrate that the toolbox introduced in the previous section provides a scalable and experimentally viable ground state preparation strategy harnessing the capabilities of PR simulators. To this end, we benchmark our approach across several target Hamiltonians $H_{\mathrm{tar}}$, including spin-$1/2$ models (Section~\ref{subsec:spin12}), free fermionic models (Section~\ref{subsec:fermion}), and spin-$1$ models (Section~\ref{subsec:spin1}). As figures of merit for the benchmarks, we consider the residual energy density,
\begin{equation}
    \varepsilon = \frac{\left|E\left(\bm{\Theta}_{\mathrm{opt}}\right) - E_{\mathrm{ex}}\right|}{N_2},
\end{equation}
and the infidelity density,
\begin{equation}
    \tilde{\mathcal{I}} = 1 - \big(\left|\braket{\psi\left(\bm{\Theta}_{\mathrm{opt}}\right) | \psi_{\mathrm{exact}}}\right|\big)^{1/N_2}\,,
\end{equation}
with $N_2$ being the size of the final extrapolated system (this quantity is the intensive analog of the usual infidelity $\mathcal{I}=1 - \left|\braket{\psi\left(\bm{\Theta}_{\mathrm{opt}}\right) | \psi_{\mathrm{exact}}}\right|$~\cite{Zhou2008a}). Here, $\ket{\psi\left(\bm{\Theta}_{\mathrm{opt}}\right)}$ denotes the state prepared by the optimized parametrized circuit, and $\ket{\psi_{\mathrm{exact}}}$ is the exact ground state, either obtained exactly or via Density Matrix Renormalization Group methods~\cite{schollwock11a,Orus2019}, which are effectively exact for the system sizes considered. The corresponding energies of these states are given by $E\left(\bm{\Theta}_{\mathrm{opt}}\right) = \braket{\psi\left(\bm{\Theta}_{\mathrm{opt}}\right)|H_{\mathrm{tar}}|\psi\left(\bm{\Theta}_{\mathrm{opt}}\right)}$ and $E_{\mathrm{ex}} = \braket{\psi_{\mathrm{ex}}|H_{\mathrm{tar}}|\psi_{\mathrm{ex}}}$, respectively.

\subsection{Spin-$1/2$ models~\label{subsec:spin12}}

\begin{figure}[tb]
    \centering
    \includegraphics[width=\linewidth]{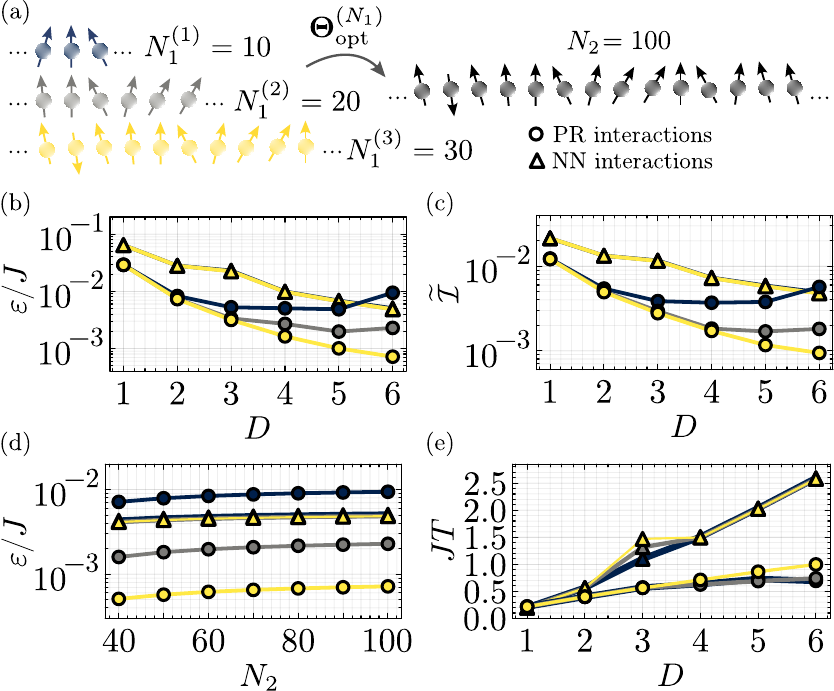}
    \caption{(a) Scheme of the protocol used to obtain the optimal parameters, $\bm{\Theta}_{\mathrm{opt}}^{(N_1)}$, for approximating the ground state of the TFIM. We optimize the circuits in easier to simulate systems with $N_{1}=10$ (dark blue), $20$ (light gray) and $30$ spins (yellow), and then use these parameters to prepare the ground state of a $N_2=100$ spins (dark gray). (b-c) Residual energy and infidelity per site, respectively, as a function of the circuit depth for different training sizes $N_1$ and the same target size $N_2$, comparing exponentially-decaying PR (circles) and NN (triangles) interactions. (d) Residual energies for a $D=6$ layers circuit as a function of the system size $N_2$ for the different circuits considered. (e) Total physical time found after the optimization (in units of the inverse of the interaction, $J^{-1}$) needed to implement the circuits shown in panels (b-c).}
    \label{fig:scaling}
\end{figure}
The first target model we consider is a prototypical benchmark for quantum spin systems, that is, the transverse-field Ising model (TFIM),
\begin{equation}~\label{eq:H_TFIM}
    H_{\mathrm{TFIM}} = -J\sum_{i}\sigma_i^x\sigma_{i+1}^x - h\sum_i \sigma_i^z\,.
\end{equation}
This model undergoes a quantum phase transition at zero temperature at the critical point $|J/h|=1$. The ground state at this point is precisely the one we target, since its long-range correlations between spins make it the hardest to prepare for this model~\cite{Sachdev2011}.

In Fig.~\ref{fig:scaling}(a), we first schematically illustrate the pre-compiled extrapolation strategy used for this TFIM, where we use training sizes $N_1=10$, $20$ and $30$ to finally extrapolate the results to a system with $N_2=100$ spins. All the circuits alternate separated quenches under the Ising interaction Hamiltonian (either the exponentially-decaying $H_{\text{PR}}^{(\text{Ising})}$ or its NN version) and the on-site driving $H_{\text{L}}=h\sum_i \sigma_{i}^{z}$, with each pair of these quenches defining a single layer. Figs.~\ref{fig:scaling}(b) and~\ref{fig:scaling}(c) display the residual energy density and infidelity, respectively, for a target system size $N_2=100$ as a function of the number of layers $D$, and for different training sizes $N_1$ (color-coded). Circle markers correspond to circuits using the PR Ising resource Hamiltonians [Eq.~\eqref{eq:H_exp_dec_ising}], while triangle markers are used for the NN resource. The results show that PR circuits consistently approximate the target ground state more accurately than their NN counterparts while requiring fewer layers. This confirms that the advantage suggested in Refs.~\cite{Tabares2023,Lyu2023} persists for substantially larger system sizes. The dependence on the training size becomes relevant only for sufficiently large circuit depths, where larger $N_1$ improves performance at fixed $N_2$. This behavior is further confirmed in Fig.~\ref{fig:scaling}(d), where we fix $D=6$ layers and plot the residual energy density as a function of the target size $N_2$ for different training sizes $N_1$. Furthermore, the slow growth of the residual energy with increasing $N_2$ of this figure demonstrates the scalability of the pre-compilation procedure. Finally, Fig.~\ref{fig:scaling}(e) shows the total interaction time (in units of $J^{-1}$) required to implement the circuits of Figs.~\ref{fig:scaling}(b) and~\ref{fig:scaling}(c), obtained by summing the interaction times of all layers. There, we observe that PR circuits not only achieve lower residual energies and infidelities than their NN counterparts, but also require shorter total interaction times. This reduction is particularly relevant in experiments, where dissipation and decoherence accumulate over time and degrade the fidelity of the final state.

\begin{figure}[tb]
    \centering
    \includegraphics[width=\linewidth]{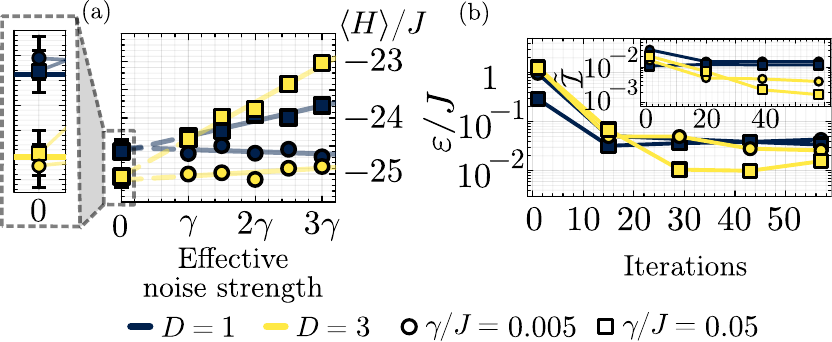}
    \caption{(a) ZNE calculation to mitigate the effects of noise and estimate $\langle H_\mathrm{TFIM}\rangle$, for a $N_1=20$ and $g/J=1$. Considering $M=10000$ shots per observable and two different values of the noise $\gamma/J$, the energy can be measured for different effective noise strengths, enabling an extrapolation to the zero noise limit. Inset: zoom of the extrapolation, with error bars as the standard deviation of the fit and solid lines representing each noiseless value. (b) Quantum-classical optimization using the optimal parameters found for a $N_{1}=16$ TFIM, deployed into a $N_2=36$ TFIM. To model an experiment, we use a global depolarizing channel of the different amplitudes considered in panel (a), and assume $M=10000$ shots per function evaluation. A calibration error of amplitude $\sigma=0.05$ [see Eq.~\eqref{eq:noise_params}] is considered for each parameter. ZNE is applied to mitigate the noise at each energy evaluation, according to the linear extrapolation and the data samples shown in (a). The main panel shows the residual energy, and the inset the infidelity.}
    \label{fig:noise}
\end{figure}

To further test the advantage of PR quantum circuits under dissipation, in Fig.~\ref{fig:noise} we now consider the effect of noise and imperfect experimental control. These effects are mitigated applying the hybrid re-optimization strategies and ZNE scheme introduced in Section~\ref{subsec:zero}. Our goal is to assess whether reliable state preparation remains possible under realistic conditions. We begin in Fig.~\ref{fig:noise}(a) by illustrating the ZNE procedure for the target TFIM with $N_1=20$ spins assuming still a perfect control of the resource Hamiltonian parameters. The optimal parameters $\bm{\Theta}_{\mathrm{opt}}$, obtained in the noiseless case for $D=1$ and $D=3$ layers (blue and yellow, respectively), are used to evaluate the energy $\langle H_{\mathrm{TFIM}}\rangle$ in the presence of global depolarizing noise applied after each interaction layer. This type of channel is a simple yet good approximation for modelling the effects of decoherence in analog quantum dynamics~\cite{Tran2023_measuringanalog,Vovrosh_2021,Andersen2025}. For each infinitesimal timestep in the quantum evolution under a certain unitary $U$, this channel considers that the quantum state $\rho$ is mapped onto the maximally depolarized state with probability $\gamma$, whereas the evolution without errors, $U\rho U^{\dagger}$, happens with probability $1-\gamma$~\cite{ding2024robustgroundstateenergyestimation}. We consider two noise amplitudes, $\gamma/J = 5\times10^{-2}$ and $5\times10^{-3}$ (circles and squares, respectively), acting throughout all the interaction time. The markers show the measured energies for different effective noise strengths proportional to $\gamma$, i.e., $\{1,3/2,2,5/2,3\}\times \gamma/J$, while the solid lines represent linear fits used for the extrapolation. The dashed lines indicate the extrapolated $\gamma\rightarrow0$ values. The inset zooms into the small-noise region, showing that the extrapolated estimates approach closely the noiseless energies (horizontal solid lines), particularly for the larger noise case $\gamma/J=5\times10^{-2}$.

After validating the ZNE procedure, we incorporate it into the full iterative protocol. Following the programming toolbox of Fig.~\ref{fig:intro}, we first perform classical optimization of the TFIM energy at the critical point for iterative training sizes up to $N_1=16$, using exponentially decaying long-range circuits with $D=1$ and $D=3$. The optimized parameters $\bm{\Theta}_{\mathrm{opt}}^{(N_1)}$ are then used as initial conditions for a re-optimization in a larger system size $N_2=36$. Each cost-function evaluation in this step is calculated assuming $M=10000$ measurement shots and evaluated at the effective noise strength values shown in Fig.~\ref{fig:noise}(a), that allow the extrapolation to the zero noise limit shown in the inset (with the solid lines representing the exact, noiseless values). This error mitigation pipeline is implemented in each step of the re-optimization, which we show in Fig.~\ref{fig:noise}(b). There, we plot the residual energy as a function of the iteration steps in the optimization of this larger system size, simulated under realistic values of $\gamma$ conditions and considering Gaussian perturbations with a variance $\sigma=0.05$ (that model a calibration uncertainty of $5\%$). Although these perturbations degrade the prepared state fidelities initially, the pre-compilation provides a favorable starting point for hybrid re-optimization on the noisy device. From these results, we conclude that the combination of noise mitigation and hardware-level re-optimization allows the residual energy and infidelity to converge close to their ideal values, with final deviations limited by the shot-noise scale, i.e., in this case $1/\sqrt{M}\sim 10^{-2}$. The infidelity density shown in the inset of Fig.~\ref{fig:noise}(b) further demonstrates the accuracy of the prepared state, yielding results around $\tilde{\mathcal{I}}\sim 2.5\times 10^{-3}$. Overall, this demonstrates that the scalable pre-compilation strategy remains effective even under noise and calibration imperfections.

Having verified that noise mitigation and hybrid re-optimization effectively recover near-ideal performance, in the following benchmarks we restrict to the noiseless scenario. This allows us to assess the intrinsic scalability of the pre-compilation strategy without relating it with hardware-specific noise effects.

\subsection{Fermionic models~\label{subsec:fermion}}

The next target model we consider is the fermionic Kitaev chain~\cite{Kitaev2001}, whose Hamiltonian reads
\begin{align}~\label{eq:H_kitaev}
\begin{split}
H_{\mathrm{Kitaev}} = &\sum_i\left[-t\left(c^{\dagger}_{i+1} c_i +\mathrm{H.c.}\right)
+ \Delta \left(c^{\dagger}_{i+1} c^{\dagger}_{i}+\mathrm{H.c.}\right)\right] \\
&- \mu\sum_i\left(c^{\dagger}_i c_i - \frac{1}{2}\right)\,.
\end{split}
\end{align}

This model is of particular interest for three main reasons. First, using the Jordan-Wigner transformation, the model maps onto the TFIM of Eq.~\eqref{eq:H_TFIM} when $t=\Delta=J$ and $\mu=2h$~\cite{Surace2022a}, and therefore provides access to the same challenging, long-range correlated ground states. Second, it can be studied using fermionic resource Hamiltonians, which connects with the growing interest in fermionic quantum processors~\cite{OBrien2018,Gonzalez-Cuadra_2023,Arguello-Luengo2019,Arguello-Luengo2020,Arguello-Luengo2021,Arguello_Luengo2022,Gkritsis2024,Li2026arxiv,Tabares2025,cavallar2025phasesensitivemeasurementsfermihubbardquantum}, where fermionic Hamiltonians can be implemented directly using analog or hybrid digital-analog approaches. Third, and more importantly for this work, when using the quadratic PR fermionic Hamiltonians of Eq.~\eqref{eq:H_resource_kitaev} as the resource, the circuit dynamics remain exactly solvable. This allows us to explore several regimes hard to access with the previous spin resource Hamiltonians, namely, studying larger system sizes without being limited by the entanglement growth. For instance, we will be able to compare the exponential and power-law decays in the interaction profile $f_L(|i-j|)$, that would require prohibitive bond dimensions in cases with strong long-range interactions. On top of that, we can efficiently compute the full evolutions generated by the simultaneous action of the on-site and interaction terms within a single layer, as defined in Eq.~\eqref{eq:gate_set}. Since these are the evolutions more naturally occurring in analog simulators, its comparison with the quenched or Trotterized versions considered for the TFIM of the previous section is also relevant for current experiments. Let us finally emphasize that, although the target Hamiltonian in this fermionic case is the same (up to a Jordan-Wigner transformation) than the one in Section~\ref{subsec:spin12}, the fermionic resource $H_{\text{PR}}^{(\text{Kitaev})}$ cannot be mapped to $H_{\text{PR}}^{(\text{I})}$, establishing a key difference between both studies.

\begin{figure}[tb]
    \centering
    \includegraphics[width=\linewidth]{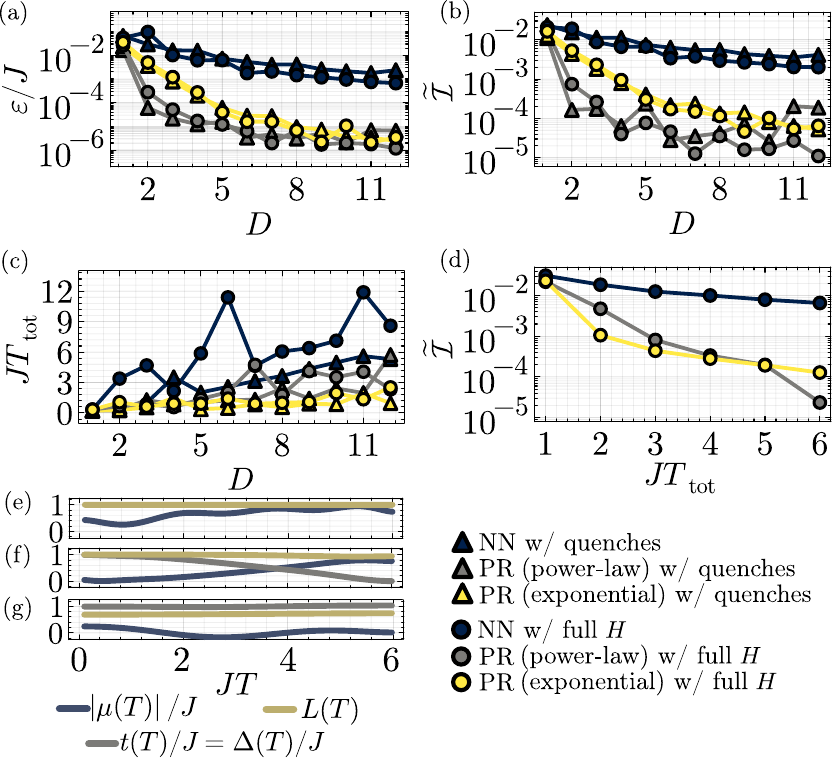}
    \caption{(a-b) Residual energies and infidelities per site, respectively, in a Kitaev chain with $N_2=1000$ modes after obtaining the optimal parameters in a $N_1=200$ one. We compare circuits built with NN (dark blue) and PR with exponential (light gray) and power-law (yellow) decay. We also depict the results obtained with a full Hamiltonian evolution (triangle) and the ones obtained with individual quenches (circles). (c) Total interaction time of the optimal circuits in panels (a-b), in units of the inverse interaction parameter $J^{-1}$. (d) Infidelity per site in a $N_2=1000$ modes Kitaev chain after obtaining the optimal parameters using quantum optimal control for different total pulse times $JT_\mathrm{tot}$. (e-g) Optimal pulses used to prepare the states shown in panel (d) for a total pulse time, $JT_{\mathrm{tot}}=6$, for the NN, PR with exponential decay, PR with power-law decay, respectively.}
    \label{fig:scaling_kitaev}
\end{figure}

In Figs.~\ref{fig:scaling_kitaev}(a) and~(b), we show the residual energy density and infidelity for a target Kitaev chain with $N_2=1000$ modes, using an iterative pre-compilation strategy with training sizes ranging from $N_1=10$ to $N_1=200$. Different colors correspond to distinct resource Hamiltonians: NN (dark blue), PR exponential (yellow), and PR power-law (gray). Markers distinguish between circuits constructed from independent quenches of the on-site and interaction terms to define a layer (triangles) and those implementing the full evolution simultaneously (circles). We observe no significant difference between these two constructions, which is experimentally relevant since the latter is simpler to implement. Moreover, PR circuits clearly outperform the NN architecture even at these large system sizes, with power-law interactions providing the best performance at shallow depths. At larger depths all circuits saturate, indicating that further improvements require increasing the training size $N_1$. Fig.~\ref{fig:scaling_kitaev}(d) shows the total interaction time as a function of circuit depth. While the interaction time of NN circuits grows approximately linearly with $D$, the PR circuits exhibit an apparent saturation around $JT_{\mathrm{tot}}\sim 2$. This demonstrates that, beyond a certain interaction time, the main bottleneck becomes the number of optimization parameters rather than the physical evolution time itself.

To further investigate this behavior, we adopt a quantum optimal control framework~\cite{Khaneja2005,Falamarzi2010,Glasser2015,Caneva2009}, promoting the layer-dependent parameters to continuous time-dependent functions, $\{t(D),\Delta(D),\mu(D),L(D)\} \rightarrow \{t(T),\Delta(T),\mu(T),L(T)\}$, which we optimize using a variant of the Gradient Ascent Pulse Engineering (GRAPE) algorithm~\cite{Khaneja2005}. The total interaction time $T_{\mathrm{tot}}$ is discretized in steps $\Delta T=0.1J^{-1}$, and the parameters within each time step are treated as constant. Exploiting the classical simulability of the quadratic model, we directly maximize the fidelity with respect to the exact ground state and include a penalty term on pulse derivatives to enforce smoothness. The resulting infidelity density is shown in Fig.~\ref{fig:scaling_kitaev}(d) as a function of $T_{\mathrm{tot}}$, confirming that PR circuits outperform NN architectures by orders of magnitude also in this continuous time optimization. The optimized control pulses for NN, exponential, and power-law interactions are displayed in Figs.~\ref{fig:scaling_kitaev}(e)--(g), respectively, for a total pulse duration $JT_{\mathrm{tot}}=6$. These pulses are significantly smoother than the standard layer-wise quenches typically considered in VQE approaches, while retaining comparable accuracy, so they provide a more experimentally practical route for preparing many-body states.

\subsection{Spin-$1$ models~\label{subsec:spin1}}

\begin{figure}[tb]
    \centering
    \includegraphics[width=\linewidth]{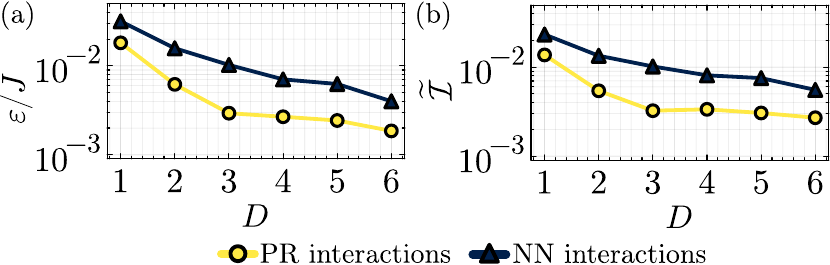}
    \caption{(a-b) Residual energy [(a)] and infidelity [(b)] per site as a function of the circuit depth, comparing PR and NN interactions for the spin$-1$ BC model with $h/J=1.1$ and $D/J=0.3135$. The system is pre-optimized for training sizes $N_1\in\{10,15,20,25\}$ and the results shown are evaluated for $N_2 = 50$ spins.}
    \label{fig:spin1}
\end{figure}

To further generalize our conclusions, we consider a spin-$1$ extension of the TFIM, namely the BC spin chain~\cite{Blume66,Capel1966,Blume1971}. The Hamiltonian for this model reads
\begin{equation}~\label{eq:BC_Hamiltonian}
    H_{\mathrm{BC}} = -J\sum_{i} S_i^x S_{i+1}^x 
    + D\sum_{i} (S_i^{x})^2 
    + h\sum_{i} S_i^z\,,
\end{equation}
where $S_i^j$ denote the spin-$1$ operators acting on site $i$. This model exhibits ferromagnetic and paramagnetic phases separated by both continuous and first-order transition lines, which meet at a tri-critical point~\cite{Alcaraz1985,Qiu1986,Getelina2024}. For benchmarking purposes, we focus on the point $(h/J,D/J)=(1.1,0.3135)$, located along the critical line where the transition belongs to the same universality class as the TFIM~\cite{Xavier2011}.

We compare circuits constructed from the spin-$1$ Ising resource Hamiltonian of Eq.~\eqref{eq:Ising_LR_spin1}, considering both the NN and PR-exponential versions. Each of the layers is implemented via three quenches: one generated by the interaction term, followed by a second quench under a global rotation generated with $\sum_i S_{i}^{z}$, and a third one given by the evolution generated with $\sum_i (S_{i}^{x})^2$. These circuits are the immediate generalization of the ones considered in Section~\ref{subsec:spin12}, but now for the spin-$1$ case and including the anisotropy. We apply the iterative pre-compilation strategy using training sizes $N_1 \in \{10,15,20,25\}$, and evaluate the resulting circuits for a target size $N_2=50$ spins. The results of this optimization protocol are shown in Figs.~\ref{fig:spin1}(a) and~\ref{fig:spin1}(b), where we plot the residual energy density and infidelity as functions of the circuit depth $D$. As in the spin-$1/2$ case, PR circuits (yellow triangles) systematically outperform their NN counterparts (dark blue triangles), particularly at shallow depths. This highlights the advantage of  PR interactions for preparing accurate ground states with reduced circuit depth also in systems with larger local Hilbert-space dimension than standard spin-$1/2$ models.

\section{Application: Non-Equilibrium Dynamics and Thermalization~\label{sec:dynamics}}

\begin{figure}[tb]
    \centering
    \includegraphics[width=\linewidth]{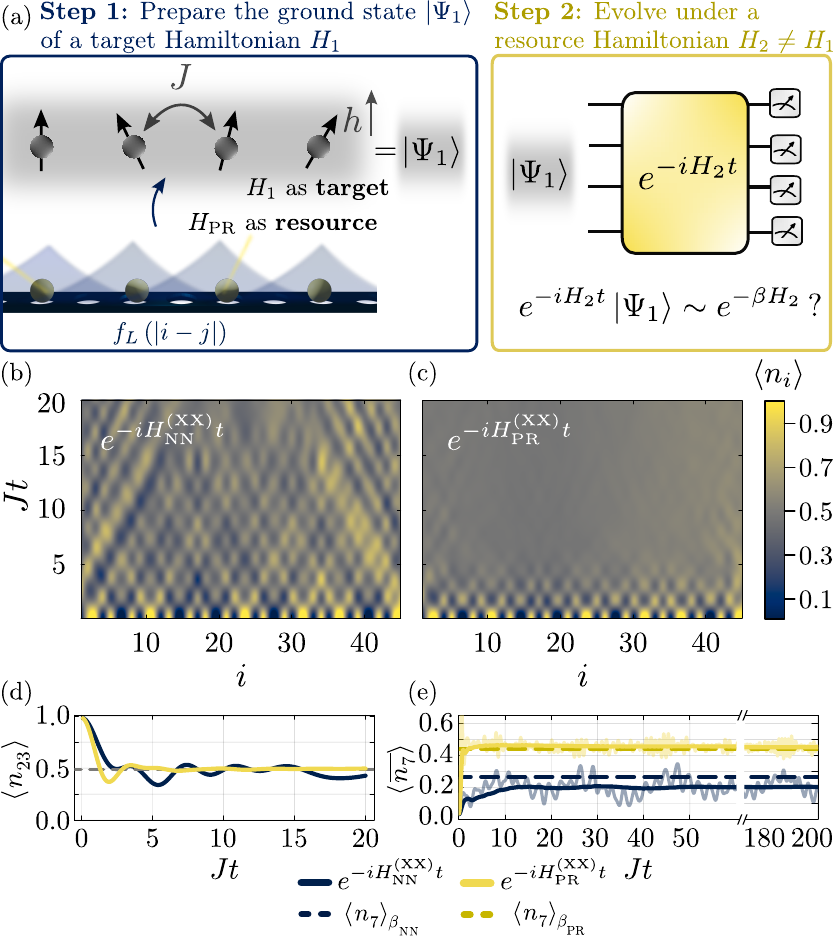}
    \caption{(a) Experimental protocol: the ground state $\ket{\Psi_1}$ of a target Hamiltonian $H_1$ is prepared using the PR toolbox, and subsequently quenched under a different Hamiltonian $H_2$ to study its dynamics. (b-c) Time evolution of site occupations $n_i$ after quenching the ground state of the AAH model ($\alpha=(\sqrt{5}-1)/2$, $h/J=4$, $N=45$), under NN [(b)] and PR [(c)] XX Hamiltonians with $L = -1/\log(0.6)$. The initial state is prepared with a $D=2$ circuit of the form Eq.~\eqref{eq:AAH_circuit}. (d) Dynamics of the central site occupation ($i=23$), with the dashed line indicating the initial filling fraction. (e) Exact diagonalization results ($N=13$) for a quenched initial state of the AAH model with $\alpha=(\sqrt{5}-1)/2$, $h/J=2$, showing long-time averages and dynamics of $n_7$ after a quench with NN and PR XX Hamiltonians with $h/J = 0.5$, $L=-1/\log(0.8)$. and weak integrability-breaking $g/J=0.01$. Dashed lines denote thermal expectation values at the corresponding effective temperatures $J\beta_{\mathrm{NN}}\approx1.84$ and $J\beta_{\mathrm{PR}}\approx0.31$.} 
    \label{fig:dynamics}
\end{figure}

As a final application of PR simulators and our state-preparation toolbox, we demonstrate how they can be used to experimentally probe signatures of the (non-)thermalization transition~\cite{Neyenhuis2017,Sugimoto2022,Mattes2025}, using the exponentially-decaying PR as the main resource. The proposed experiment is schematically depicted in Fig.~\ref{fig:dynamics}(a), and consists of two steps. First, we prepare the ground state $\ket{\Psi_1}$ of a target Hamiltonian $H_1$, and then we perturb this state by evolving it with a different resource Hamiltonian, $H_2$. Since $\ket{\Psi_1}$ is not be an eigenstate of $H_2$, it will undergo a non-trivial dynamical evolution given by:
\begin{equation}
    \ket{\Psi(t)} = e^{-i H_2 t}\ket{\Psi_1}\,.
\end{equation}
If $H_2$ is integrable, the dynamics preserve the memory of the initial state and thus is not expected to thermalize. On the contrary, if $H_2$ is a non-integrable model, the model is expected to thermalize. A typical way of assessing thermalization consists of looking at the long-time average of certain observable $O$,
\begin{equation}~\label{eq:long_time_average}
   \langle \bar{O} \rangle
   =
   \lim_{t\rightarrow\infty}
   \frac{1}{t}
   \int_0^t
   \langle \Psi(\tau)| O |\Psi(\tau)\rangle\, d\tau,
\end{equation}
and checking whether it converges to the thermal expectation value
\begin{equation}
    \langle \bar{O}\rangle\rightarrow \langle O\rangle_\beta
    =
    \mathrm{Tr}(O\rho_\beta),
    \qquad
    \rho_\beta
    =
    \frac{e^{-\beta H_2}}
    {\mathrm{Tr}(e^{-\beta H_2})},
\end{equation}
where $\beta$ is fixed by canonical equation,
\begin{equation}~\label{eq:canonical_equation}
   \langle \Psi_1|H_2|\Psi_1\rangle
   =
   \mathrm{Tr}(H_2\rho_\beta).
\end{equation}

Here, we illustrate this protocol by considering the Aubry-Andr\'e-Harper (AAH) model~\cite{Harper1955,Aubry1980,Thouless1983} as the initial target Hamiltonian $H_1$. This Hamiltonian reads:
\begin{equation}~\label{eq:H_AAH}
    H_{\mathrm{AAH}} = -J\sum_{i}\left(\sigma_{i}^{+}\sigma_{i+1}^- + \mathrm{H.c.} \right)
    - h\sum_j \cos\!\left(2\pi\alpha j\right)\sigma_j^z\,,
\end{equation}
which corresponds to the NN hopping Hamiltonian with a quasi-periodic potential. For irrational $\alpha$, the model exhibits a localization transition at $h/J=2$, separating extended ($h/J<2$) and localized ($h/J>2$) phases~\cite{Harper1955,Aubry1980,Thouless1983}. The choice of this model allows us to demonstrate the PR toolbox with an additional example while still having an integrable model with a simple interpretation of its phases. As such, for the state preparation we target the ground state in the localized regime, choosing $\alpha=(\sqrt{5}-1)/2$, $h/J=4$, and $N_2=45$ spins. 

As initial state $\ket{\Psi_0}$, we take a product configuration aligned with the sign of the local field. We find that a shallow circuit with $D=2$ layers,
\begin{equation}~\label{eq:AAH_circuit}
    \ket{\Psi_1}
    =
    \prod_{i=1}^{D}
    e^{-i H^{\mathrm{PR}}_{\mathrm{AAH}}(J=1,L_i,h_i)T_i}
    \ket{\Psi_0},
\end{equation}
prepares the target ground state with an infidelity density below $10^{-3}$. Note that the resource Hamiltonian $H^{\mathrm{PR}}_{\mathrm{AAH}}$ used in the circuit of Eq.~\eqref{eq:AAH_circuit} is the PR version of the AAH Hamiltonian of Eq.~\eqref{eq:H_AAH}, replacing the NN part by the PR exponential Hamiltonian of Eq.~\eqref{eq:H_exp_dec_xx}.

In the second step, we evolve this state with both the NN and PR exponential resource Hamiltonians of Eq.~\eqref{eq:H_exp_dec_xx}, that we denote here by $H_{\mathrm{NN}}^{(\text{XX})}$ and $H_{\mathrm{PR}}^{(\text{XX})}$, respectively. While $H_{\mathrm{NN}}^{(\text{XX})}$ is integrable, the long-range interactions of $H_{\mathrm{PR}}^{(\text{XX})}$ break integrability and, thus, are expected to induce thermalization. This highlights the interaction range as a tunable control parameter to probe many-body relaxation. Fig.~\ref{fig:dynamics}(b) and~(c) show the evolution of the local excitation density $n_i=(1+\sigma_i^z)/2$ under NN and PR quenches, respectively. In both cases, the initially localized excitations spread across the chain. However, the NN dynamics display ballistic propagation and persistent coherent oscillations, whereas the PR dynamics rapidly approach a spatially uniform distribution, erasing memory of the initial configuration. This contrast is further illustrated in Fig.~\ref{fig:dynamics}(d), where the middle-site occupation $n_{23}(t)$ equilibrates only under $H_{\mathrm{PR}}^{(\text{XX})}$, approaching the  expected mean value.

To provide a quantitative comparison with thermal predictions, we repeat the protocol for a smaller chain of $N_2=13$, which can be solved exactly. In this case, we prepare the ground state at $h/J=2$ and quench to $h/J=0.5$, considering both NN and PR evolutions. We also include a weak transverse field $V=g\sum_i \sigma_i^x$ with $g/J=0.01$, modeling parasitic fields of similar magnitude that may appear in real experiments. Although $V$ breaks magnetization conservation (and hence full integrability), it remains perturbative on timescales $t\ll O(1/g^2)$~\cite{Kollar2011}. This means that, in these timescales, the system can be considered as integrable in practice. Furthermore, thanks to the exact solution, we can now use Eq.~\eqref{eq:canonical_equation} to determine the effective temperature $\beta$ of the states, and compute the corresponding thermal expectation values. As shown in Fig.~\ref{fig:dynamics}(e), the PR evolution (in solid yellow lines) converges toward the thermal prediction (dashed lines), while the (quasi-)integrable NN dynamics (in dark blue) do not, even for long times $Jt\sim 200$. 

Overall, these results demonstrate that PR interactions provide a powerful experimental knob to control integrability and probe many-body thermalization within currently accessible system sizes.

\section{Conclusions~\label{sec:conclusions}}

Summing up, we have introduced a scalable and experimentally viable state-preparation toolbox for  programmable long-range analog quantum simulators. By combining classical pre-compilation, noise-mitigation techniques, and hardware-level re-optimization, we demonstrate that programmable-range interactions significantly enhance both the spread of correlations and the trainability of parametrized circuits across spin-$1/2$, fermionic, and spin-$1$ target models. Our results show that these advantages persist up to system sizes of $10^2$-$10^3$ particles, while remaining compatible with realistic noise levels and calibration imperfections. Beyond equilibrium state preparation, we illustrate how dynamical control over the interaction range provides a powerful experimental knob to probe many-body thermalization and integrability breaking in system sizes which are state-of-the-art in PR analog quantum simulators. Altogether, our findings establish programmable long-range simulators as a versatile platform for exploring complex quantum many-body phenomena at scales beyond classical reach. In future works, we plan to apply our state-preparation toolbox beyond the one-dimensional models considered in this work, e.g., to higher-dimensional target~\cite{Semeghini2021,Liu2022,Qin2022a} and resource Hamiltonians~\cite{douglas15a,Gonzalez-Tudela2015b,Hung2016,Gonzalez-Tudela2018,Gonzalez-Tudela2019a,Arguello-Luengo2019,Arguello-Luengo2021,Arguello_Luengo2022}, which would further test its scalability beyond classically tractable regimes. 

\begin{acknowledgements}
  The authors acknowledge useful discussions with Luca Tagliacozzo and Diego Porras during the development of this work. The authors also acknowledge support from the CSIC Research Platform on Quantum Technologies PTI-001. C.T. acknowledges support from Comunidad de Madrid (PIPF-2022/TEC-25625) and also from Fundaci\'on Humanismo y Ciencia. AMH acknowledges support from Fundaci\'on General CSIC's ComFuturo program, which has received funding from the European Union's Horizon 2020 research and innovation program under the Marie Sk{\l}odowska-Curie grant agreement No. 101034263. AGT, JTS and CTL acknowledge support from Spanish project Proyecto PID2024-162384NB-I00 financed by MICIU/AEI/10.13039/501100011033 and from FEDER,UE. AGT and JTS also acknowledge support from the Programa Fundamentos FBBVA through the grant EIC24-1-17304. AGT acknowledges support from the QUANTERA project MOLAR with reference PCI2024153449 and funded MICIU/AEI/10.13039/501100011033 and by the European Union.  JTS acknowledges the support by the Ministry for Digital Transformation and of Civil Service of the Spanish Government through the QUANTUM ENIA project call - Quantum Spain project and by the European Union through the Recovery, Transformation and Resilience Plan - NextGenerationEU within the framework of the Digital Spain 2026 Agenda.

\end{acknowledgements}
\bibliographystyle{apsrev4-2}
\bibliography{referencesAlexNew,references}
\end{document}